\documentclass[twocolumn]{aastex631}

\usepackage{xcolor}
\usepackage{soul}

\graphicspath{{./}{figures/}}

\shorttitle{TeV Halo Candidate Surrounding Radio-quiet pulsar}
\begin{document}

\title{HAWC Detection of a TeV Halo Candidate Surrounding a Radio-quiet pulsar}

\correspondingauthor{Sara Coutiño de León}
\email{scoutino@icecube.wisc.edu}
\correspondingauthor{Ke Fang}
\email{kefang@physics.wisc.edu}

\author[0000-0003-0 197-5646]{A.~Albert}
\affiliation{Physics Division, Los Alamos National Laboratory, Los Alamos, NM, USA }
\author[0000-0001-8749-1647]{R.~Alfaro}
\affiliation{Instituto de F\'{i}sica, Universidad Nacional Autónoma de México, Ciudad de Mexico, Mexico }
\author{J.C.~Arteaga-Velázquez}
\affiliation{Universidad Michoacana de San Nicolás de Hidalgo, Morelia, Mexico }
\author[0000-0003-3207-105X]{E.~Belmont-Moreno}
\affiliation{Instituto de F\'{i}sica, Universidad Nacional Autónoma de México, Ciudad de Mexico, Mexico }
\author[0000-0003-2158-2292]{T.~Capistrán}
\affiliation{Instituto de Astronom\'{i}a, Universidad Nacional Autónoma de México, Ciudad de Mexico, Mexico }
\author[0000-0002-8553-3302]{A.~Carramiñana}
\affiliation{Instituto Nacional de Astrof\'{i}sica, Óptica y Electrónica, Puebla, Mexico }
\author[0000-0002-6144-9122]{S.~Casanova}
\affiliation{Institute of Nuclear Physics Polish Academy of Sciences, PL-31342 IFJ-PAN, Krakow, Poland }
\author[0000-0002-1132-871X]{J.~Cotzomi}
\affiliation{Facultad de Ciencias F\'{i}sico Matemáticas, Benemérita Universidad Autónoma de Puebla, Puebla, Mexico }
\author[0000-0002-7747-754X]{S.~Coutiño de León}
\affiliation{Department of Physics, University of Wisconsin-Madison, Madison, WI, USA }
\author[0000-0001-9643-4134]{E.~De la Fuente}
\affiliation{Departamento de F\'{i}sica, Centro Universitario de Ciencias Exactase Ingenierias, Universidad de Guadalajara, Guadalajara, Mexico }
\author{R.~Diaz Hernandez}
\affiliation{Instituto Nacional de Astrof\'{i}sica, Óptica y Electrónica, Puebla, Mexico }
\author[0000-0002-2987-9691]{M.A.~DuVernois}
\affiliation{Department of Physics, University of Wisconsin-Madison, Madison, WI, USA }
\author[0000-0002-0087-0693]{J.C.~Díaz-Vélez}
\affiliation{Departamento de F\'{i}sica, Centro Universitario de Ciencias Exactase Ingenierias, Universidad de Guadalajara, Guadalajara, Mexico }
\author[0000-0001-7074-1726]{C.~Espinoza}
\affiliation{Instituto de F\'{i}sica, Universidad Nacional Autónoma de México, Ciudad de Mexico, Mexico }
\author{K.L.~Fan}
\affiliation{Department of Physics, University of Maryland, College Park, MD, USA }
\author[0000-0002-0173-6453]{N.~Fraija}
\affiliation{Instituto de Astronom\'{i}a, Universidad Nacional Autónoma de México, Ciudad de Mexico, Mexico }
\author[0000-0002-5387-8138]{K.~Fang}
\affiliation{Department of Physics, University of Wisconsin-Madison, Madison, WI, USA }
\author[0000-0002-4188-5584]{J.A.~García-González}
\affiliation{Tecnologico de Monterrey, Escuela de Ingenier\'{i}a y Ciencias, Ave. Eugenio Garza Sada 2501, Monterrey, N.L., Mexico, 64849}
\author[0000-0003-1122-4168]{F.~Garfias}
\affiliation{Instituto de Astronom\'{i}a, Universidad Nacional Autónoma de México, Ciudad de Mexico, Mexico }
\author[0000-0002-6738-9351]{Armelle Jardin-Blicq}
\affiliation{Université Bordeaux, CNRS/IN2P3, LP2I Bordeaux, UMR 5797, F-33170 Gradignan, France}
\affiliation{Max-Planck Institute for Nuclear Physics, D-69117 Heidelberg, Germany}
\author[0000-0002-5209-5641]{M.M.~González}
\affiliation{Instituto de Astronom\'{i}a, Universidad Nacional Autónoma de México, Ciudad de Mexico, Mexico }
\author[0000-0002-9790-1299]{J.A.~Goodman}
\affiliation{Department of Physics, University of Maryland, College Park, MD, USA }
\author[0000-0001-9844-2648]{J.P.~Harding}
\affiliation{Physics Division, Los Alamos National Laboratory, Los Alamos, NM, USA }
\author[0000-0002-2565-8365]{S.~Hernandez}
\affiliation{Instituto de F\'{i}sica, Universidad Nacional Autónoma de México, Ciudad de Mexico, Mexico }
\author[0000-0002-3808-4639]{D.~Huang}
\affiliation{Department of Physics, Michigan Technological University, Houghton, MI, USA }
\author[0000-0002-5527-7141]{F.~Hueyotl-Zahuantitla}
\affiliation{Universidad Autónoma de Chiapas, Tuxtla Gutiérrez, Chiapas, México}
\author[0000-0001-5811-5167]{A.~Iriarte}
\affiliation{Instituto de Astronom\'{i}a, Universidad Nacional Autónoma de México, Ciudad de Mexico, Mexico }
\author[0000-0003-4467-3621]{V.~Joshi}
\affiliation{Erlangen Centre for Astroparticle Physics, Friedrich-Alexander-Universit\"at Erlangen-N\"urnberg, Erlangen, Germany}
\author[0000-0001-6336-5291]{A.~Lara}
\affiliation{Instituto de Geof\'{i}sica, Universidad Nacional Autónoma de México, Ciudad de Mexico, Mexico }
\author[0000-0002-2153-1519]{J.~Lee}
\affiliation{University of Seoul, Seoul, Rep.of Korea }
\author[0000-0001-5516-4975]{H.~León Vargas}
\affiliation{Instituto de F\'{i}sica, Universidad Nacional Autónoma de México, Ciudad de Mexico, Mexico }
\author[0000-0003-2696-947X]{J.T.~Linnemann}
\affiliation{Department of Physics and Astronomy, Michigan State University, East Lansing, MI, USA}
\author[0000-0001-8825-3624]{A.L.~Longinotti}
\affiliation{Instituto de Astronom\'{i}a, Universidad Nacional Autónoma de México, Ciudad de Mexico, Mexico }
\author[0000-0003-2810-4867]{G.~Luis-Raya}
\affiliation{Universidad Politecnica de Pachuca, Pachuca, Hgo, Mexico }
\author[0000-0001-8088-400X]{K.~Malone}
\affiliation{Space Science and Applications Group, Los Alamos National Laboratory, Los Alamos, NM, USA }
\author[0000-0001-9052-856X]{O.~Martinez}
\affiliation{Facultad de Ciencias F\'{i}sico Matemáticas, Benemérita Universidad Autónoma de Puebla, Puebla, Mexico }
\author[0000-0002-2824-3544]{J.~Martínez-Castro}
\affiliation{Centro de Investigaci\'on en Computaci\'on, Instituto Polit\'ecnico Nacional, M\'exico City, M\'exico.}
\author[0000-0002-2610-863X]{J.A.~Matthews}
\affiliation{Dept of Physics and Astronomy, University of New Mexico, Albuquerque, NM, USA }
\author[0000-0001-9361-0147]{J.A.~Morales-Soto}
\affiliation{Universidad Michoacana de San Nicolás de Hidalgo, Morelia, Mexico }
\author[0000-0002-1114-2640]{E.~Moreno}
\affiliation{Facultad de Ciencias F\'{i}sico Matemáticas, Benemérita Universidad Autónoma de Puebla, Puebla, Mexico }
\author[0000-0002-7675-4656]{M.~Mostafá}
\affiliation{Department of Physics, Pennsylvania State University, University Park, PA, USA }
\author[0000-0003-0587-4324]{A.~Nayerhoda}
\affiliation{Institute of Nuclear Physics Polish Academy of Sciences, PL-31342 IFJ-PAN, Krakow, Poland }
\author[0000-0003-1059-8731]{L.~Nellen}
\affiliation{Instituto de Ciencias Nucleares, Universidad Nacional Autónoma de Mexico, Ciudad de Mexico, Mexico }
\author[0000-0001-9428-7572]{M.~Newbold}
\affiliation{Department of Physics and Astronomy, University of Utah, Salt Lake City, UT, USA }
\author[0000-0002-6859-3944]{M.U.~Nisa}
\affiliation{Department of Physics and Astronomy, Michigan State University, East Lansing, MI, USA }
\affiliation{Department of Physics, Michigan Technological University, Houghton, MI, USA}
\author[0000-0002-8774-8147]{Y.~Pérez Araujo}
\affiliation{Instituto de Astronom\'{i}a, Universidad Nacional Autónoma de México, Ciudad de Mexico, Mexico }
\author[0000-0001-5998-4938]{E.G.~Pérez-Pérez}
\affiliation{Universidad Politecnica de Pachuca, Pachuca, Hgo, Mexico }
\author[0000-0002-6524-9769]{C.D.~Rho}
\affiliation{University of Seoul, Seoul, Rep. of Korea}
\author[0000-0003-1327-0838]{D.~Rosa-González}
\affiliation{Instituto Nacional de Astrof\'{i}sica, Óptica y Electrónica, Puebla, Mexico }
\author[0000-0001-8644-4734]{M.~Schneider}
\affiliation{Department of Physics, University of Maryland, College Park, MD, USA }
\author{J.~Serna-Franco}
\affiliation{Instituto de F\'{i}sica, Universidad Nacional Autónoma de México, Ciudad de Mexico, Mexico }
\author[0000-0002-1012-0431]{A.J.~Smith}
\affiliation{Department of Physics, University of Maryland, College Park, MD, USA }
\author{Y.~Son}
\affiliation{University of Seoul, Seoul, Rep.of Korea }
\author[0000-0002-1492-0380]{R.W.~Springer}
\affiliation{Department of Physics and Astronomy, University of Utah, Salt Lake City, UT, USA }
\author[0000-0001-9725-1479]{K.~Tollefson}
\affiliation{Department of Physics and Astronomy, Michigan State University, East Lansing, MI, USA }
\author[0000-0002-1689-3945]{I.~Torres}
\affiliation{Instituto Nacional de Astrof\'{i}sica, Óptica y Electrónica, Puebla, Mexico }
\author{R.~Torres-Escobedo}
\affiliation{Tsung-Dao Lee Institute, Shanghai Jiao Tong University, Shanghai, China}
\author{X.~Wang}
\affiliation{Department of Physics, Michigan Technological University, Houghton, MI, USA }
\author{K.~Whitaker}
\affiliation{Department of Physics, Pennsylvania State University, University Park, PA, USA }
\author[0000-0002-6623-0277]{E.~Willox}
\affiliation{Department of Physics, University of Maryland, College Park, MD, USA }
\author[0000-0003-0513-3841]{H.~Zhou}
\affiliation{Tsung-Dao Lee Institute, Shanghai Jiao Tong University, Shanghai, China}
\author[0000-0002-8528-9573]{C.~de León}
\affiliation{Universidad Michoacana de San Nicolás de Hidalgo, Morelia, Mexico }
\collaboration{80}{(THE HAWC COLLABORATION)}

\begin{abstract}
Extended very-high-energy (VHE; 0.1-100~TeV) $\gamma$-ray emission has been observed around several middle-aged pulsars and referred to as ``TeV halos". Their formation mechanism remains under debate. It is also unknown whether they are ubiquitous or related to certain subgroup of pulsars. With 2321 days of observation, the High Altitude Water Cherenkov (HAWC) Gamma-Ray Observatory detected VHE $\gamma$-ray emission at the location of the radio-quiet pulsar PSR J0359+5414 with $>6\sigma$ significance. By performing likelihood tests with different spectral and spatial models and comparing the TeV spectrum with multi-wavelength observations of nearby sources, we show that this excess is consistent with a TeV halo associated with PSR J0359+5414, though future observation of HAWC and multi-wavelength follow-ups are needed to confirm this nature.  This new halo candidate is located in a non-crowded region in the outer Galaxy. It shares similar properties to the other halos but its pulsar is younger and radio-quiet. Our observation implies that TeV halos could commonly exist around pulsars and their formation does not depend on the configuration of the pulsar magnetosphere. 
\end{abstract}

\section{Introduction}\label{sec:intro}
Extended TeV gamma-ray emission has been observed around several middle-aged ($>100$~kyr) pulsars and grouped as a new source class named ``TeV halos" \citep{2017PhRvD..96j3016L, 2022NatAs...6..199L}. Seven sources are referenced as TeV halos in the online catalog for TeV Astronomy, TeVCat\footnote{\url{http://tevcat2.uchicago.edu/}} \citep{tevcat}, including the first halos around the Geminga and Monogem pulsars, discovered by HAWC \citep{2017Sci...358..911A},  HESS J1825-137 reported by the H.E.S.S. collaboration \citep{2018A&A...612A...2H}, and the halo of PSR J0622+3749, identified by the LHAASO collaboration \citep{2021PhRvL.126x1103A}. The VHE fluxes of these halos suggest that $\sim 10\%-40\%$ of the spin-down power of the pulsars is converted into $e^{\pm}$ pair population that interacts with the ambient interstellar radiation field \citep{2019PhRvD.100d3016S, 2021PhRvL.126x1103A}. The diffusion coefficients derived from the sizes of the halos are typically two orders of magnitude lower than the average diffusion coefficient of the interstellar medium (ISM; \citealp{PhysRevD.96.103013, 2019PhRvD.100d3016S}).  

The formation mechanism of the TeV halos is still under debate \citep{2017PhRvD..96j3016L, 2019PhRvD.100d3016S,Giacinti:2019nbu,2022NatAs...6..199L, 2022arXiv220704011L, 2022arXiv220508544D}. Whether they are related to the local environment, such as extended, diffuse emission by other sources near the pulsar  (e.g., the Monogem Ring \citealp{1996ApJ...463..224P}), is also questioned. If TeV halos commonly exist around pulsars, they can be used to study the propagation of cosmic rays (e.g., \citealp{Evoli:2018aza}) and to identify pulsars that are otherwise invisible to radio and $\gamma$-ray observations \citep{2017PhRvD..96j3016L}. 

In this letter, we report the detection of a new TeV halo candidate around the pulsar PSR J0359+5414 (hereafter J0359) using 2321 days of HAWC data. The detection of J0359 was first reported in the {\it Fermi} Large Area Telescope (LAT) First Source Catalog (1FGL, \citealp{1fgl}) where it remained as an unclassified source until the Third Source Catalog (3FGL, \citealp{3fgl}). J0359 was later classified as a radio-quiet pulsar by \cite{2017ApJ...834..106C} with an age of 75 kyr and a spin-down power of $\dot{E}=1.3\times 10^{36}\,\mbox{erg}\,\mbox{s}^{-1}$. In \cite{2018MNRAS.476.2177Z} a pseudo-distance of J0359 is reported as $d=3.45$~kpc, derived from the $\dot{E}$ and the gamma-ray flux. The latest report at high energies of J0359 appears in the {\it Fermi}-LAT Fourth Source Catalog (4FGL, \citealp{4fgl}) where it is detected above $33\sigma$ in the MeV-GeV energy range. A pulsar wind nebula (PWN) with an extension of $\sim 30 ''$ was observed by {\it Chandra} as a result of a X-ray analysis on gamma-ray pulsars \citep{2018MNRAS.476.2177Z}. No radio emission has been detected from the pulsar \citep{2021A&A...654A..43G}. The VHE $\gamma$-ray emission from the vicinity of J0359 observed by HAWC presents similar properties to the other TeV halos candidates, including the derived acceleration efficiency and diffusion coefficient. If this source is a TeV halo, it would support the hypothesis that the halos are ubiquitous. 

The paper is organized as follows. The data set and analysis framework are described in Section~\ref{sec:data}. The results of the spectral and spatial analysis are presented in  Section~\ref{sec:res}. In Section~\ref{sec:discu}, the broadband spectral energy distribution (SED) of J0359 is presented and the origin of the TeV emission is discussed. The conclusions are summarized in Section~\ref{sec:conc}.

\begin{figure}[t]
\plotone{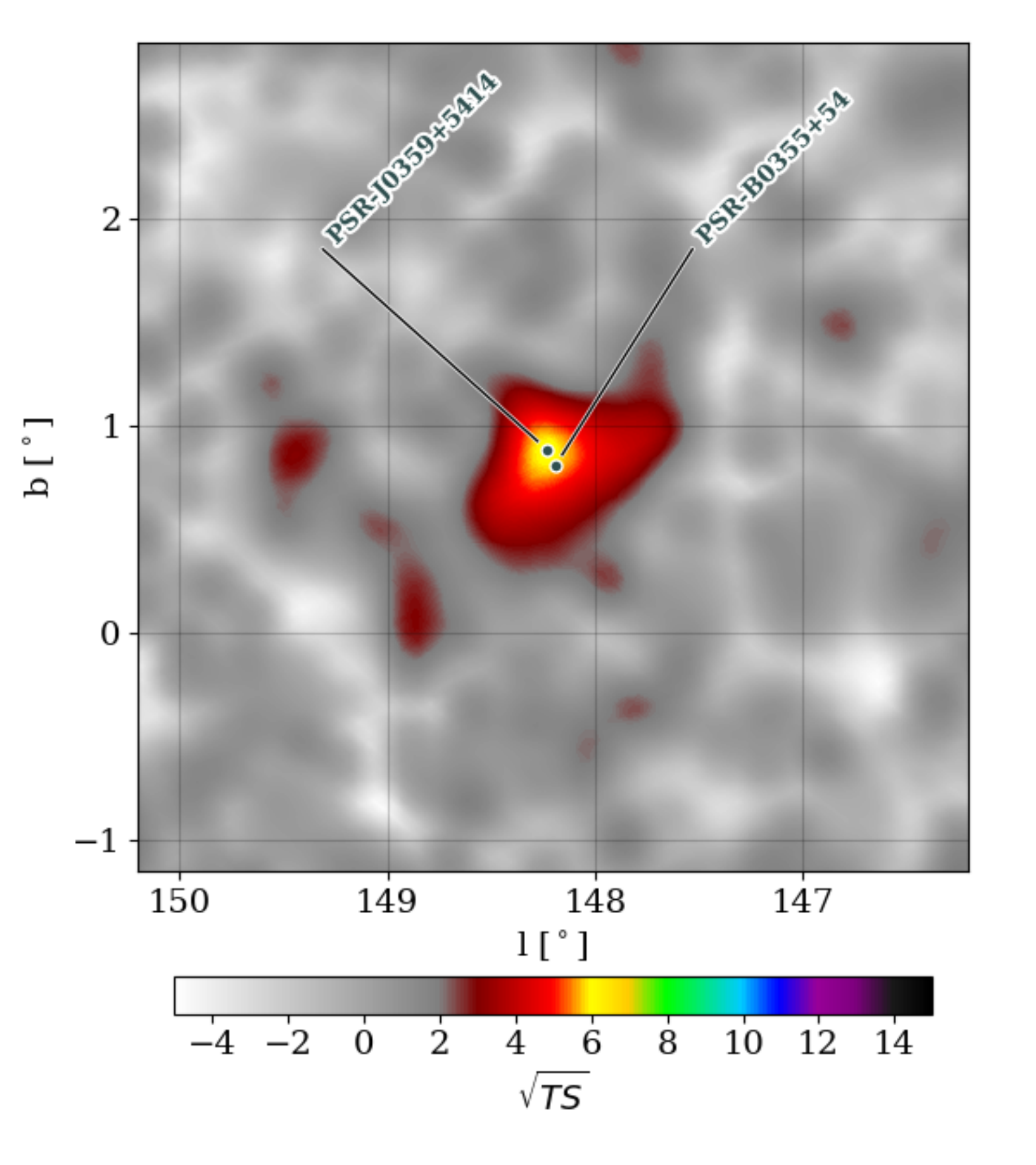}
\caption{HAWC significance map in Galactic coordinates using 2321 days of live data. The significance is computed with a point-like spatial template and a power-law spectrum with spectral index $\alpha=2.7$. For comparison, the positions of PSR J0359+5414 and PSR B0355+54 are marked.  
\label{fig:j0359-data}}
\end{figure}

\section{Instrument and Data Analysis}\label{sec:data}

The HAWC Gamma-Ray Observatory consists of 300 water Cherenkov detectors located at $19^{\circ}$N in Puebla, Mexico at an altitude of 4100~m. Each detector is instrumented with 4 photo-multiplier tubes (PMTs) that are capable of detecting the Cherenkov radiation produced in the detector water when an electromagnetic or hadronic shower hits the ground, which is initiated by a $\gamma$-ray or a cosmic ray, respectively, when it enters the Earth's atmosphere. HAWC is sensitive to sources with declinations between $-41^{\circ}$ and $+79^{\circ}$ and to energies in the 300 GeV to $> 100$ TeV range. The data set used in this analysis comprises 2321 days of live data taken from November 2014 to October 2021. The data set is divided into 11 analysis bins ($f_{\rm Hit}$) based on the fraction of PMTs that are triggered in each event, on and off the main detector array. A full description of HAWC's design and performance can be found in \cite{2015ICRC...34..966S} and \cite{hawc2017}.

A maximum likelihood analysis was performed using the Multi-Mission Maximum Likelihood (3ML) framework \citep{3ml} with the HAWC Accelerated Likelihood (HAL) plug-in \citep{hal}.  For model selection, we use the likelihood ratio test statistic (TS) which is defined by 

\begin{equation}\label{eq:ts}
    \rm{TS} = 2\ln \frac{\mathcal{L}_{\rm S+B}}{\mathcal{L}_{\rm B}},
\end{equation}
where $\mathcal{L}_{\rm S+B}$ is the maximum likelihood of a signal plus background model, which depends on the spectral and spatial parameters, and $\mathcal{L}_{\rm B}$ is the maximum likelihood of the background-only hypothesis. Three spectral models are tested, including single power-law (PL, Equation \ref{PL}), log-parabola (LOGP, Equation \ref{LOGP}), and power-law with an exponential energy cutoff (PL+CO, Equation \ref{PL+CO}):
\begin{equation}\label{PL}
\frac{dN}{dE} = N_0 \left(\frac{E}{E_0}\right)^{-\alpha},
\end{equation}
\begin{equation}\label{LOGP}
\frac{dN}{dE} = N_0 \left(\frac{E}{E_0}\right)^{-\alpha -\beta\ln\left(E/E_0\right)},
\end{equation}
\begin{equation}\label{PL+CO}
\frac{dN}{dE} = N_0 \left(\frac{E}{E_0}\right)^{-\alpha}\times\exp\left(\frac{-E}{E_c}\right). 
\end{equation}
In the above equations, $N_0$ is the flux normalization in units of $[\mbox{TeV}^{-1}\mbox{cm}^{-2}\mbox{s}^{-1}]$, $E_0$ is the pivot energy fixed at 30 TeV to minimize correlations with the other parameters, $\alpha$ is the spectral index, $E_c$ is the cut-off energy  and $\beta$ is the curvature of the log-parabola spectrum. 
Two spatial models are tested: a point-like template and an extended template. The extended template is described by a symmetric Gaussian with width as a free parameter.

The energy range in which a source is detected is computed by multiplying a step function with the best fit model (nominal case). The lower and upper values of the step function at which the likelihood decreases by $1\sigma$, $2\sigma$ or $3\sigma$ from that of the nominal case are regarded as the upper limit to the minimum energy and lower limit to the maximum energy, respectively. 

\begin{deluxetable*}{lccccccc}
\tablecaption{PSR J0359-5414 likelihood fit results for the two spatial scenarios and different spectral shapes. \label{tab:res-part1}}
\tablewidth{0pt}
\tablehead{
\colhead{Model} & \colhead{TS} & \colhead{$\Delta{\rm BIC}$}& \colhead{Extension} & \colhead{$N_0$} & \colhead{$\alpha$} & \colhead{$\beta$} & \colhead{$E_c$} \\
\colhead{} & \colhead{} & \colhead{}&\colhead{[$\circ$]} &\colhead{[$\times 10^{-16}\mbox{TeV}^{-1}\mbox{cm}^{-2}\mbox{s}^{-1}$]} & \colhead{} & \colhead{} & \colhead{[TeV]}
}
\startdata
PL, point-like & 37.86 & -12& - & $1.34_{-0.27}^{+0.34}$ & $2.60 \pm 0.16$ & - & -\\
LOGP, point-like & 39.18 & -1&-  & $1.6_{-0.4}^{+0.5}$ & $2.80 \pm 0.23$ & $0.14 \pm 0.12$ & -\\
PL+CO, point-like & 37.98 & 0&- & $4_{-4}^{+50}$ & $2.5 \pm 1.2$ & - & $500$\\
\hline
PL, extended & 40.27 & -12 & $0.2 \pm 0.1$ & $2.0_{-0.6}^{+0.8}$ & $2.52 \pm 0.16$ & - & -\\
LOGP, extended & 41.72 & -1.2 & $0.2\pm0.1$ & $2.6_{-1.0}^{+1.5}$ & $ 2.71 \pm 0.22$ & $0.14 \pm 0.13$ & -\\
PL+CO, extended & 40.48 & 0 & $0.23 \pm 0.1$ &$14_{-4}^{+5}$ & $2.40 \pm 0.19$ & - & $270_{-130}^{+240}$\\
\enddata
\tablecomments{All the associated errors are statistical. The best model is the one with the lowest BIC value so, $\Delta{\rm BIC}$ is the difference between a model and the best model, such that it quantifies the evidence against the model with the highest BIC value. In this case, from both spatial models, the PL+CO spectral model results with the highest BIC value. The energy cutoff of 500 TeV of the PL+CO point-like model is the boundary of the fit.}
\end{deluxetable*}
\begin{figure}
    \gridline{\fig{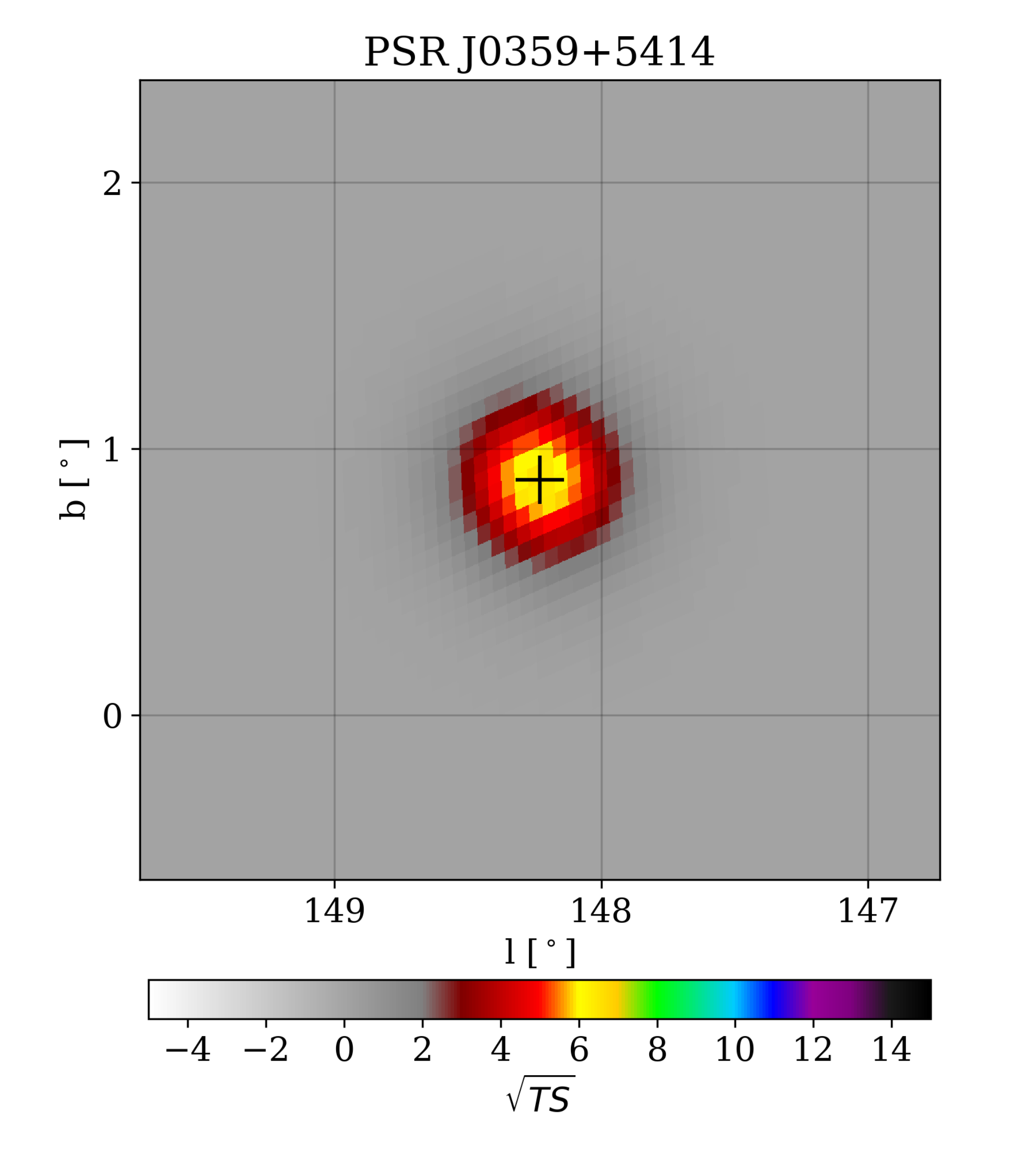}{0.25\textwidth}{(a) PL, point-like model}
              \fig{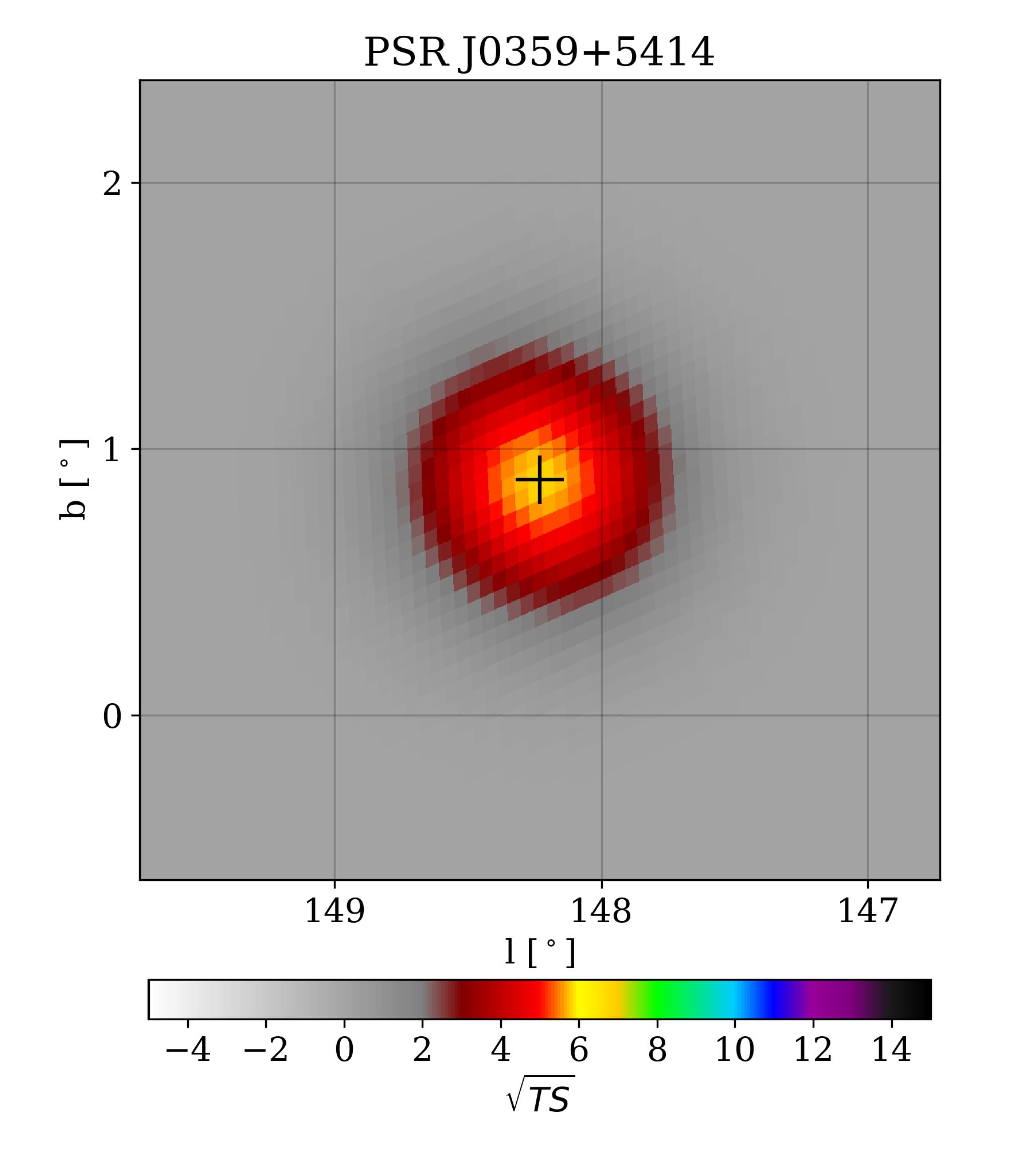}{0.25\textwidth}{(b) PL, extended model}
    }
    \gridline{\fig{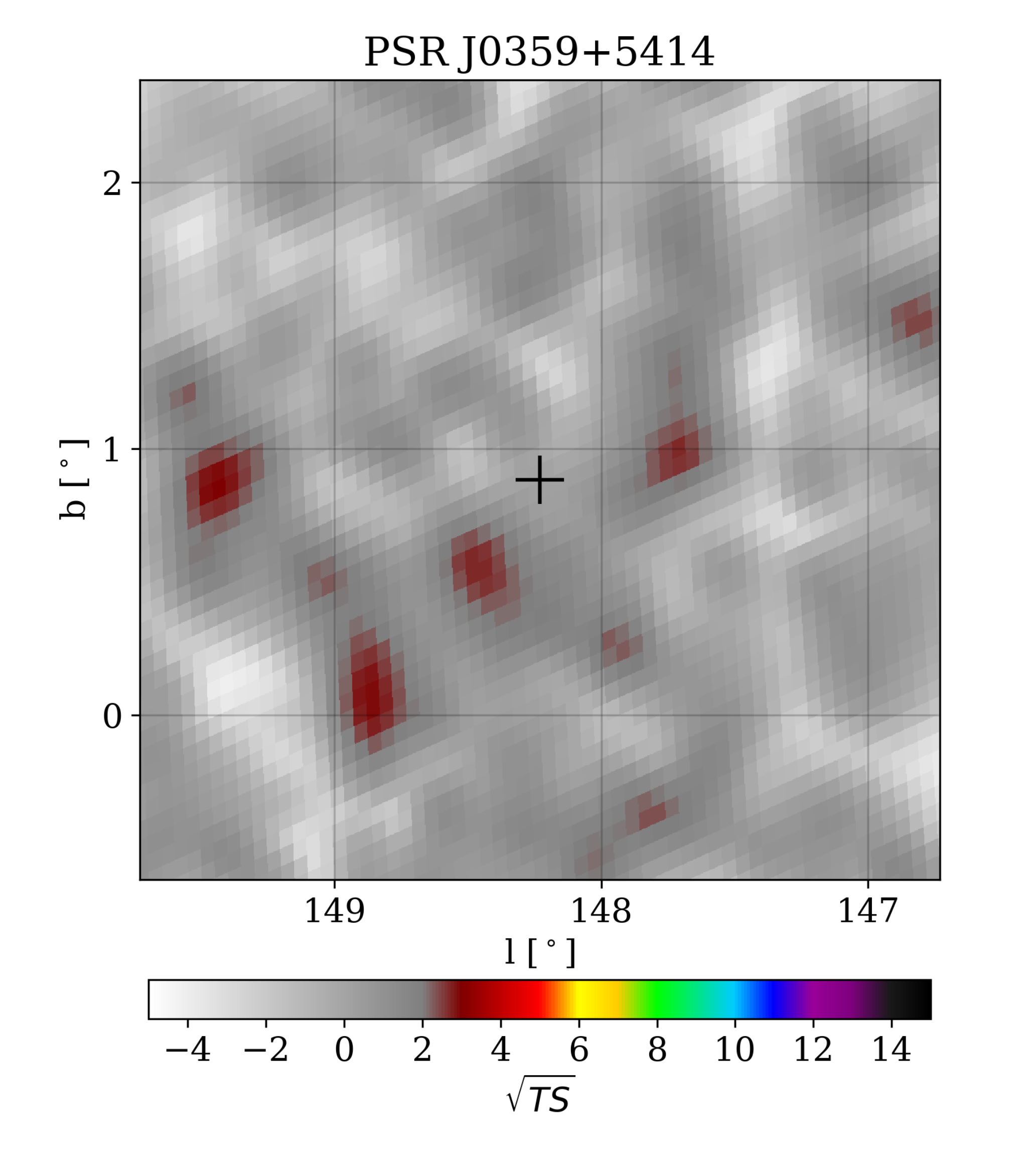}{0.25\textwidth}{(c) PL, point-like residuals}
              \fig{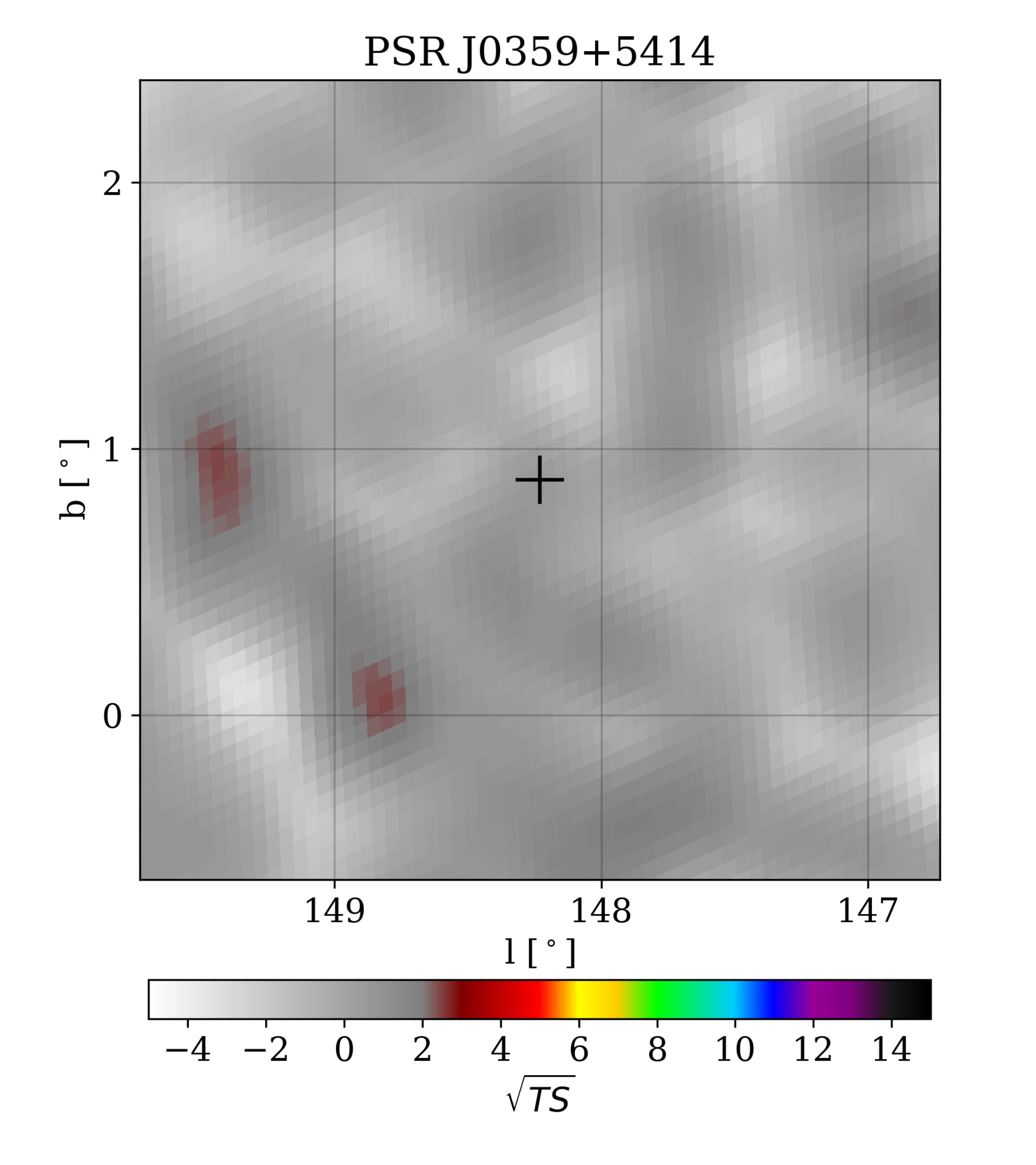}{0.25\textwidth}{(d) PL, extended residuals}
    }
    \gridline{\fig{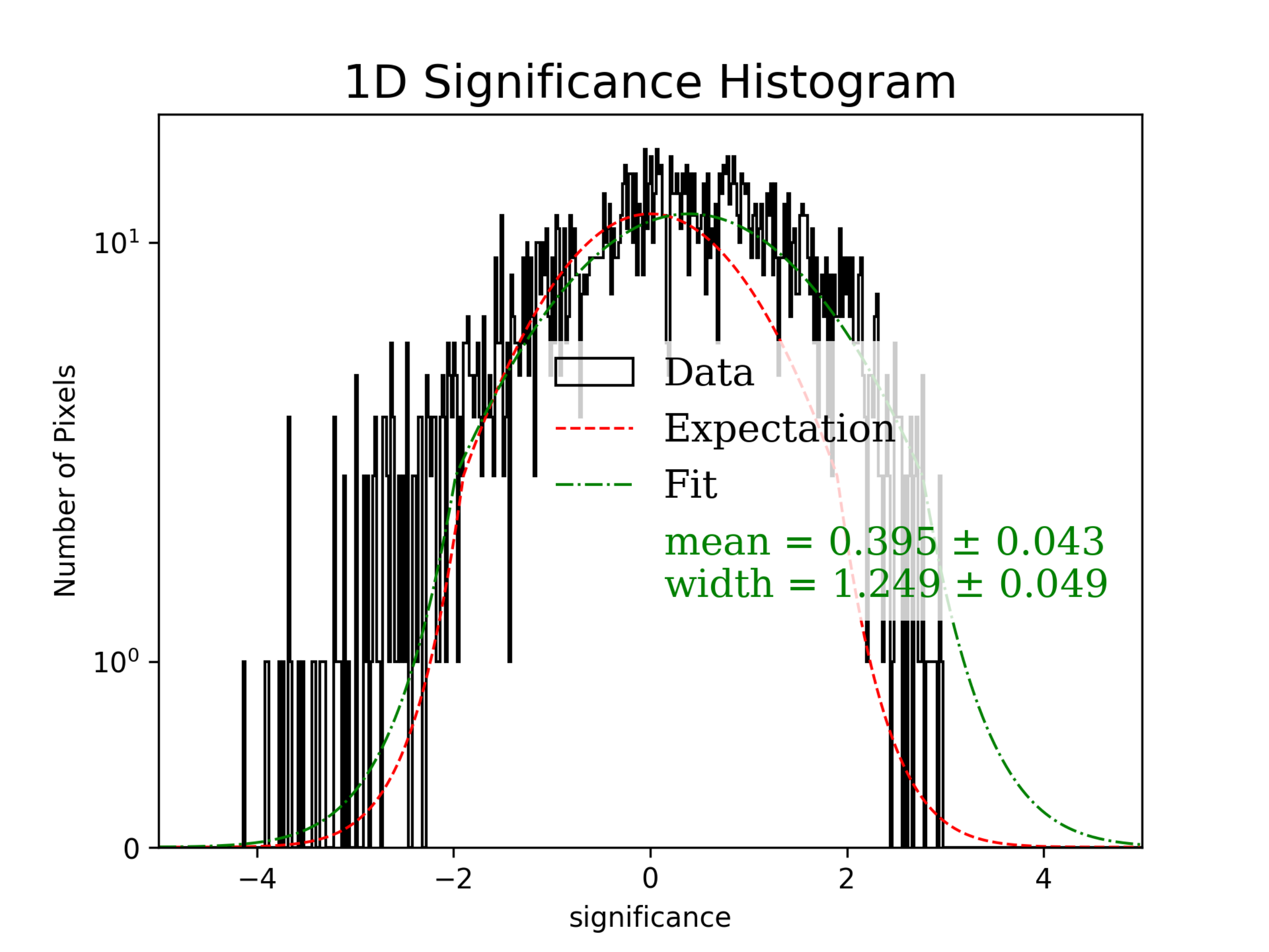}{0.25\textwidth}{(e) PL, point-like residual histogram}
              \fig{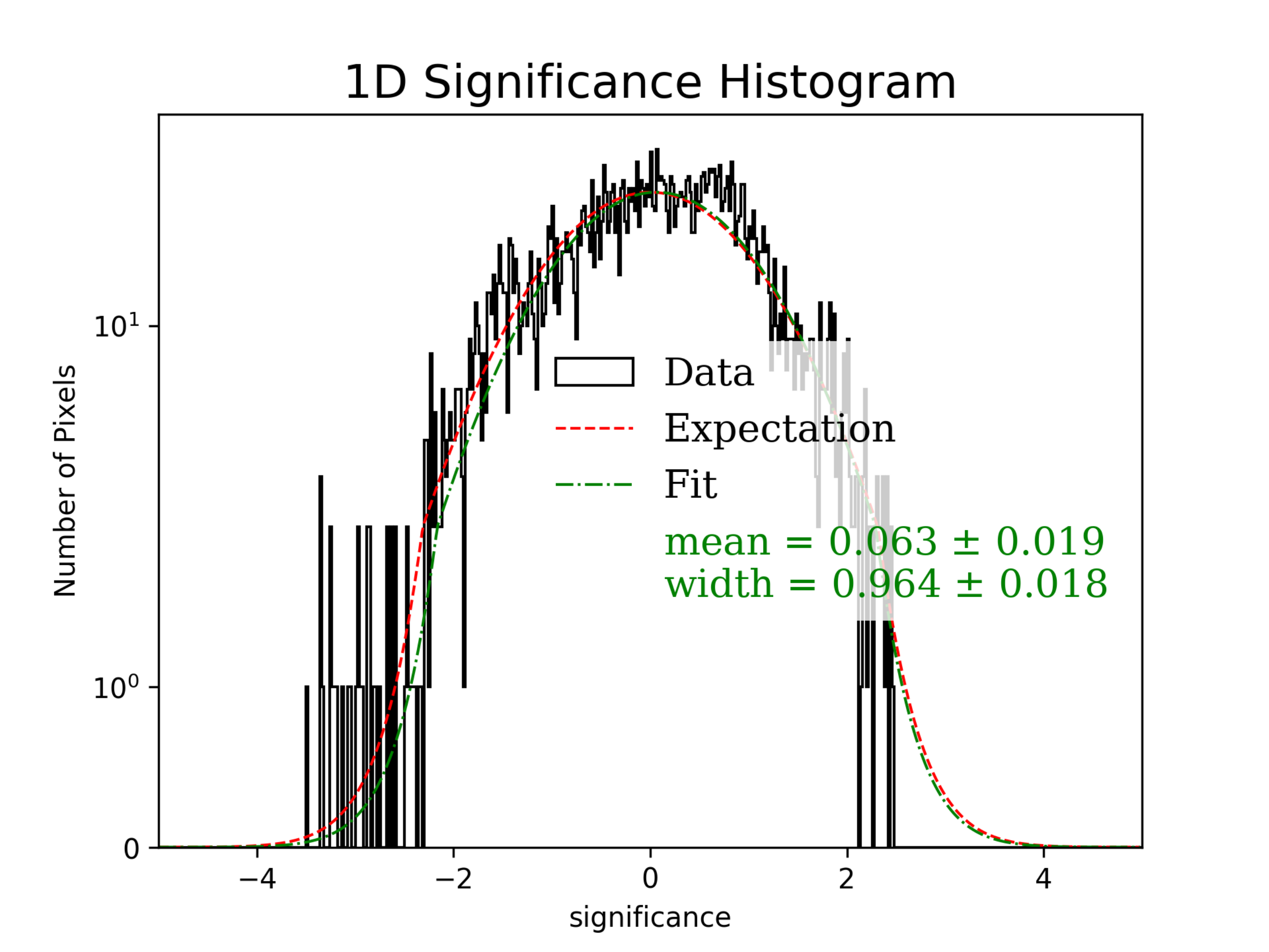}{0.25\textwidth}{( f) PL, extended residual histogram}
    }
    \caption{Comparison of the model maps, significance maps, and 1-D residual histograms for point-like and extended spatial models. The source position is fixed to PSR J0359+5414 (black cross in the significance maps) and the spectrum is assumed to be a non-broken power-law. The best-fit parameter values are listed in Table \ref{tab:res-part1}.}
    \label{fig:j0359}
\end{figure}

\section{Results}\label{sec:res}
\subsection{Association with J0359}
We first free the position of the emission and fit the PL point source model to data. The best-fit R.A. and decl. are $59.83\pm0.07_{\rm stat}$ and $54.22\pm0.05_{\rm stat}$ degrees (the systematic uncertainty at this location is $0^{\circ}.02$), which are consistent with the position of J0359 (59.86 and 54.25 degrees for R.A and decl. respectively). The TS of the model is $\rm{TS}= 38.18$, which corresponds to a significance of $ 6.18\sigma$ for four degrees of freedom based on the Wilks theorem \citep{Wilks:1938dza}. As the position is consistent with the pulsar position, we fixed the TeV emission to the pulsar position to perform the spectral analysis.

Table \ref{tab:res-part1} summarizes the best-fit parameters of different spectral and spatial models.  The simplest model assuming a point-like morphology and non-broken PL yields $\rm{TS}= 37.86$. In general more complicated models with extended morphology and spectral curvature yields a larger TS since they have more degrees of freedom than the PL point-source model. So, the preferred spectral models for both spatial assumptions is a PL, based the BIC values, where these models have the lower ones.

Figure \ref{fig:j0359} presents the model and residual significance maps, and the residual histograms for the two spatial templates assuming a PL spectral model. The residual histogram shows the distribution of the significance value in each pixel within the region of interest centered at J0359. The residual significance is defined as the deviation from the background expectation after fitting and subtracting the modeled emission from J0359. If only random background fluctuations are left, then the significance values follow a standard normal distribution (dashed red line). A positive tail is visible in the residual map of the point-source model. Although the current sample do not allow to distinguish between the different spatial models, the residual histograms in Figure \ref{fig:j0359} indicate that we get a better fit for an extended model.

The energy range of the detection are found to be 7-188 TeV at $1\sigma$ level, 11-89 TeV at $2\sigma$ level and 15-51 TeV at $3\sigma$ level, with the PL point-source model. For the PL extended model, the energy range is 4-190 TeV at $1\sigma$ level, 9-110 TeV at $2\sigma$ level and 17-78 TeV at $3\sigma$ level. 

The luminosity of the VHE emission is $L_{15-51\,\rm TeV}= 3.6\times 10^{32} \,\rm erg\,s^{-1}$ for a distance of 3.45~kpc. 
The typical energies of the synchrotron and inverse Compton photons produced by the same electrons are related by $E_{\rm syn} \approx 2.1\,\rm keV (E_{\rm IC} / 30 \,{\rm  TeV})\, (B/10\,\mu\rm G)$ (e.g., \citealp{1997MNRAS.291..162A}), where $B$ is the magnetic field strength in the PWN.  As the magnetic energy density of a PWN is usually higher than the energy density of the Cosmic Microwave Background (CMB) and infrared (IR) photons of the ISM, the synchrotron flux of a typical PWN at keV energies is expected to be higher than its inverse Compton emission at the HAWC energies (see e.g., the Crab Nebula  \citealp{2020NatAs...4..167H}). However, the X-ray luminosity of J0359's PWN, $L_{0.3-10\,{\rm keV}}=2.8\times 10^{31} \, \rm erg\,s^{-1}$ \citep{2018MNRAS.476.2177Z}, is instead $\sim 13$ times lower than the VHE gamma-ray luminosity.  This suggests the existence of a VHE electron population outside the region where the nebula is energetically dominant, which is expected in the case of a TeV halo \citep{2017PhRvD..96j3016L, 2022NatAs...6..199L}. 

\begin{figure*}[ht!]
\includegraphics[width=0.5\textwidth]{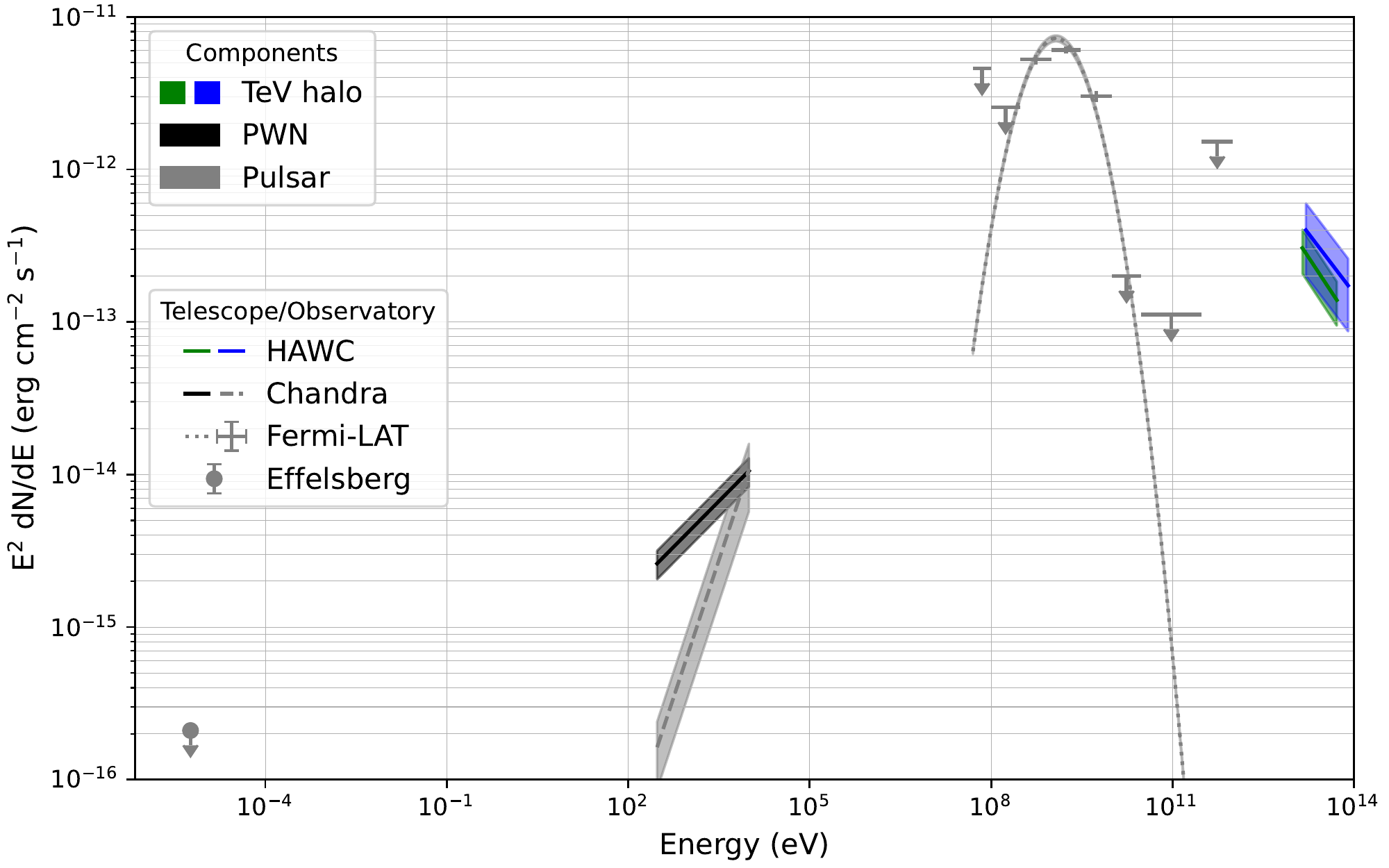}
\includegraphics[width=0.5\textwidth]{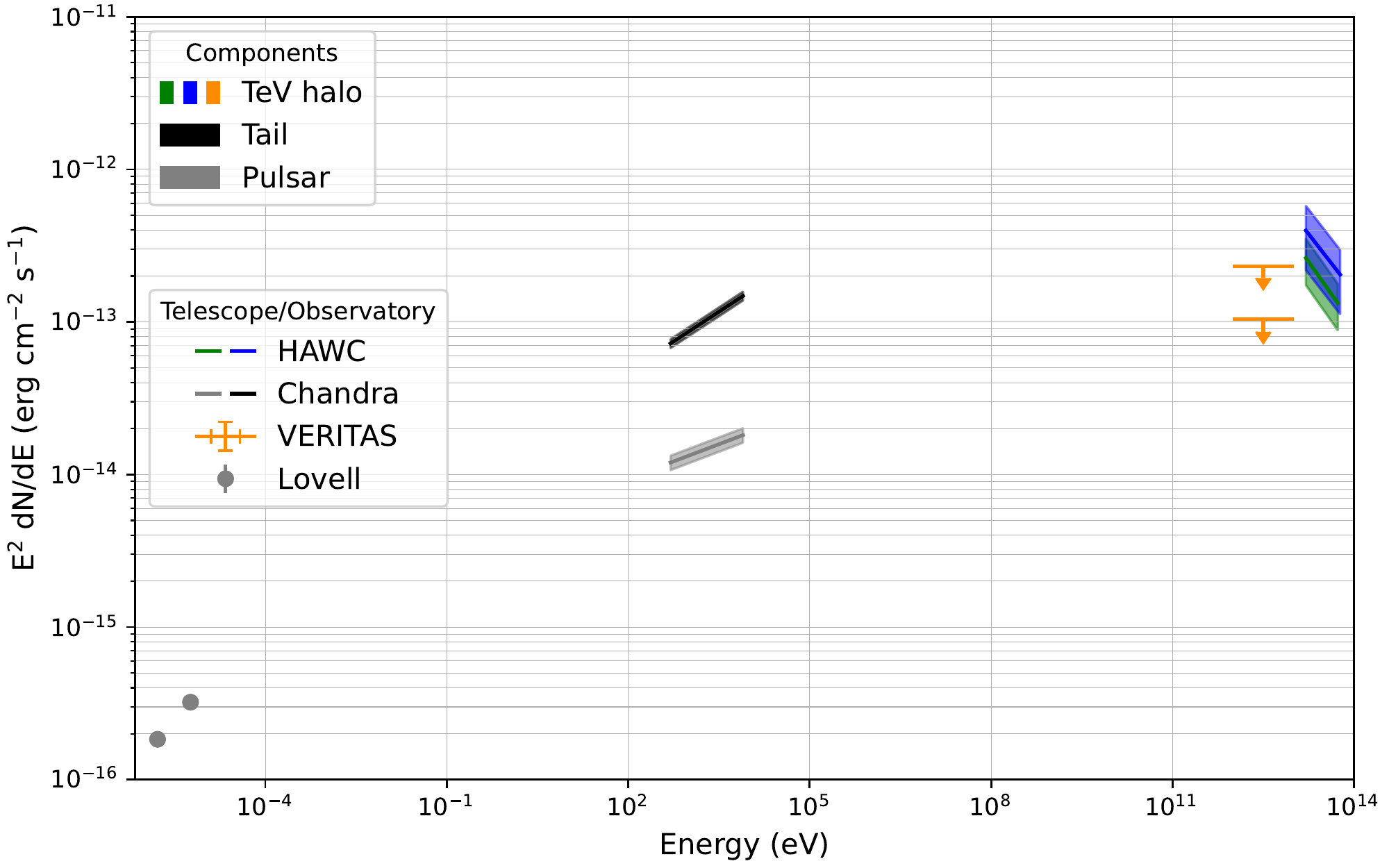}
\caption{\textbf{Left panel:} Spectral energy distribution (SED) of the emission around PSR J0359+5414, including the TeV halo (green and blue bands corresponding to the HAWC observation for a point-like and extended model, respectively, as explained in Section \ref{sec:res}), the PWN (black band at 0.3-10 keV; \citealp{2018MNRAS.476.2177Z}), and the pulsar (in grey color; including the upper limit in radio at 1400 MHz from \citealp{2021A&A...654A..43G}, the band in X-ray at 0.3-10 keV from  \citealp{2018MNRAS.476.2177Z}, and the data points or limits at 100~MeV-1~TeV from \citealp{4fgl}). \textbf{Right panel:} SED of the emission around PSR B0355+54. The green and blue bands indicate the TeV excess emission obtained from fits to the HAWC data with models that center at B0355 with point-like and extended spatial profiles, respectively (see Appendix~\ref{app:b0355}). For comparison, the  upper limits on VHE gamma-ray emission from the PWN by VERITAS with hard spectral cuts are shown in orange, with the upper and lower bars corresponding to region sizes of $0^{\circ}.1$ and $0^{\circ}.235$, respectively  \citep{2021ApJ...916..117B}. The black band at 0.5-8 keV indicates the PWN in X-rays \citep{2016ApJ...833..253K}. The grey band at 0.5-8~keV  \citep{2016ApJ...833..253K} and the circular data markers at 1400 and 1600 MHz \citep{1995MNRAS.273..411L} correspond to the emission from the pulsar. The HAWC bands correspondo to statistical uncerntanties only.
\label{fig:j0359-sed}}
\end{figure*}

Figure~\ref{fig:j0359-sed} presents the broadband SED of J0359. The pulsar, PWN, and TeV halo components are shown in grey, black, and in blue/green colors, respectively. The multi-wavelength data points include an upper limit of the pulsar emission by the Effelsberg telescope at 1400 MHz \citep{2021A&A...654A..43G}, X-ray measurements of the pulsar and PWN \citep{2018MNRAS.476.2177Z}, $\gamma$-ray observation of the pulsar from 50 MeV to 1 TeV by the {\it Fermi}-LAT \citep{4fgl}, and the VHE flux of the halo measured by HAWC. 

\subsection{Nearby pulsar B0355+54}
Another pulsar, PSR B0355+54 (B0355) is only 0.09 degrees from J0359. B0355 is
classified as a radio-loud pulsar with characteristic age of 564 kyr and spin-down power $\dot{E}=4.5\times 10^{34}\,\mbox{erg}\,\mbox{s}^{-1}$ at a distance of 1 kpc. B0355 has not been detected at high or very-high energies \citep{2021ApJ...916..117B}. Below we investigate whether B0355 is related to the HAWC excess emission. 

We performed likelihood fits and compared three scenarios: 1) the VHE emission is only associated with J0359, 2) the VHE emission is only associated with B0355, and 3) the VHE emission is contributed by both sources. We present the detailed results of scenarios 2 and 3 in Appendix \ref{app:b0355} and \ref{app:both}, respectively. We find that the two-source scenario (scenario 3) is disfavored compared to the single-source scenarios. Scenario 1 (J0359) yields lower BIC values than scenario 2 (B0355) for various spectral and spatial models, though the preference of scenario 1 is not statistically significant. 

The VERITAS telescope searched for emission from the PWN of B0355 and posed tight upper limits on the TeV flux \citep{2021ApJ...916..117B}. The right panel of   Figure \ref{fig:j0359-sed} shows the broadband SED of B0355, which includes the radio observation of the pulsar \citep{1995MNRAS.273..411L}, X-ray observation of the pulsar and its tail at  0.5-8~keV  \citep{2016ApJ...833..253K}, and the VERITAS upper limits at 95\% C.L. between 1 and 10~TeV \citep{2021ApJ...916..117B}. For comparison, we show the best-fit flux between 16 and 59~TeV obtained by assuming that the VHE emission is centered at the position of B0355. The upper limits set by VERITAS on B0355's tail are in tension with the HAWC's flux at 16 TeV for both the point-like and extended models. This suggests that the excess emission observed by HAWC is more likely associated with J0359 than B0355, though future multi-wavelength observation is needed to confirm the finding.

\section{Systematic Uncertainties}\label{sec:discu}
The systematic uncertainties arising from the detector performance and simulations are described in \cite{hawc2017} and \cite{hawc2019}. The systematic contribution is calculated in a single energy band for each spectral and spatial parameter, with the positive (negative) shift results added in quadrature to account for the upward (downward) uncertainties. 
The systematic uncertainties are calculated for the PL spectral model and for both the point-like and extended templates. 

To account for additional sources of systematic uncertainties, such as the variations in the atmosphere that are not considered in simulations, a 10\% error has been added to normalization flux \citep{3hwc}. The total systematic uncertainties are reported in Table \ref{tab:sys}. 
\begin{deluxetable}{lccc}
\tablecaption{Systematic uncertainties considering a PL for each spatial scenario. \label{tab:sys}}
\tablewidth{0pt}
\tablehead{
\colhead{Model} & \colhead{Parameter} & \colhead{Lower sys.} & \colhead{Upper sys.}\\
}
\startdata
Point-like & $N_0$ & $-3.9$ & $4.6$\\
{} & $\alpha$ & $-0.15$ & $0.3$\\
\hline
Extended & $N_0$ & $-4.6$ & $3.4$\\
{} & $\alpha$ & $-0.05$ & $0.03$\\
{} & extension & $-0.02$ & $0.02$\\
\enddata
\tablecomments{$N_0$ is in units of $10^{-17}\,\rm TeV^{-1} cm^{-2} s^{-1}$ and extension is in degrees.}
\end{deluxetable}
\section{Conclusions}\label{sec:conc}
With 2321 days of HAWC observation, VHE $\gamma$-ray emission is detected in a relatively source-empty region in the outer galaxy. Based on likelihood fits with different spectral and spatial models to the HAWC data and the comparison of VHE $\gamma$-ray flux with multi-wavelength observations, we conclude that the emission is a TeV halo candidate associated with the pulsar PSR~J0359$+$5414. 

If this TeV emission is a halo, it would share similar characteristics with the existing population. 
We find a 95\% upper limit on the extension of the emission as $ 0.^\circ41$ (with the PL-extended model in Table \ref{tab:res-part1}), corresponding to a physical size of $R_{\rm ul} = 25\,(d/3.45\,\rm kpc)\,\rm pc$. The diffusion coefficient of the halo is confined to be $D \lesssim R_{\rm ul}^2 / (4\,t_e) = 3.7\times 10^{27}\,\rm cm^2\,s^{-1}(t_e / 12\,\rm kyr)^{-1}(d/3.45\,{\rm kpc})^2$, where $t_e \sim 12\,{\rm kyr} (E_e / 100\,{\rm TeV})^{-1}$ is the cooling time of an electron at energy $E_e$ by upper-scattering the CMB. Like the other halos \citep{2017Sci...358..911A}, the diffusion coefficient is much lower than the average diffusion coefficient of the ISM. 

The candidate halo of J0359 joins the observation of extended VHE emission surrounding PSR J0622$+$3749 \citep{2021PhRvL.126x1103A} as the first evidence of TeV halos around radio-quiet pulsars. Their presence suggests that the formation of the halos is insensitive to the configuration of the pulsar magnetosphere, in particular, the geometry of the $\gamma$-ray and radio beams \citep{2001AIPC..558..115H}. 

With an age of 70~kyr, J0359 is younger than the other pulsars with halos. It is likely in a transition between the so-called relic- and halo-stage of a PWN, the boundaries of which are not well defined and have motivated different classification criteria of TeV halos \citep{2017PhRvD..96j3016L, Giacinti:2019nbu,  2022NatAs...6..199L}. Our observation of TeV halo features associated with J0359 implies that high-energy particles may already start escaping in the ISM in the late relic-stage.

Our observation provides spectral evidence toward a TeV halo nature of J0359. Future data from HAWC and multi-wavelength follow-ups of this new TeV source are crucial to confirming its nature via morphological studies that identify the halo extension and exclude the association with the nearby pulsars. Future observations of young to middle-aged pulsars like PSR~J0359$+$5414 with wide-field $\gamma$-ray experiments and imaging atmospheric Cherenkov telescopes may provide further understanding into the evolution of TeV PWNe and their connection with TeV halos.   

\section*{Acknowledgments}
We acknowledge the support from: the US National Science Foundation (NSF); the US Department of Energy Office of High-Energy Physics; the Laboratory Directed Research and Development (LDRD) program of Los Alamos National Laboratory; Consejo Nacional de Ciencia y Tecnolog\'ia (CONACyT), M\'exico, grants 271051, 232656, 260378, 179588, 254964, 258865, 243290, 132197, A1-S-46288, A1-S-22784, c\'atedras 873, 1563, 341, 323, Red HAWC, M\'exico; DGAPA-UNAM grants IG101320, IN111716-3, IN111419, IA102019, IN110621, IN110521; VIEP-BUAP; PIFI 2012, 2013, PROFOCIE 2014, 2015; the University of Wisconsin Alumni Research Foundation; the Institute of Geophysics, Planetary Physics, and Signatures at Los Alamos National Laboratory; Polish Science Centre grant, DEC-2017/27/B/ST9/02272; Coordinaci\'on de la Investigaci\'on Cient\'ifica de la Universidad Michoacana; Royal Society - Newton Advanced Fellowship 180385; Generalitat Valenciana, grant CIDEGENT/2018/034; The Program Management Unit for Human Resources \& Institutional Development, Research and Innovation, NXPO (grant number B16F630069); Coordinaci\'on General Acad\'emica e Innovaci\'on (CGAI-UdeG), PRODEP-SEP UDG-CA-499; Institute of Cosmic Ray Research (ICRR), University of Tokyo, H.F. acknowledges support by NASA under award number 80GSFC21M0002. We also acknowledge the significant contributions over many years of Stefan Westerhoff, Gaurang Yodh and Arnulfo Zepeda Dominguez, all deceased members of the HAWC collaboration. Thanks to Scott Delay, Luciano D\'iaz and Eduardo Murrieta for technical support.

\bibliography{biblio}{}
\bibliographystyle{aasjournal}

\clearpage

\appendix
\section{PSR B0355+54 fitting results}\label{app:b0355}

In this section, we explore the possibility that the TeV excess comes entirely from B0355. 
We fit models with a power-law (PL) spectrum and the spatial templates described in Section \ref{sec:res}. The results are summarized in Table \ref{tab:b0355}. The energy ranges at which the source is detected are 7-180 TeV at $1\sigma$ level, 11-90 TeV at $2\sigma$ level and 17-54 TeV assuming at $3\sigma$ level assuming a point-like morphology. For an extended morphology, the energy ranges are found to be 8-155 TeV for $1\sigma$ level, 11-90 TeV at $2\sigma$ level and 17-59 at $3\sigma$ level.

As single-source scenarios are not nested models, we have employed the Bayesian Information Criterion (BIC) to select the models. The difference in the BIC value, $\Delta {\rm BIC}$, quantifies the evidence against the model with a higher BIC value. According to \citet{bic}, if $\Delta {\rm BIC}$ is between 0 and 2 it is not clear which model is preferred; $\Delta {\rm BIC}$ between 2 and 10 and above 10 indicates a slight and strong preference of the model with the smallest BIC, respectively.

The small difference in $\Delta {\rm BIC}$ from the fits of models centered at J0359 and B0355 does not allow us to distinguish between the models. This is expected as the angular distance of the two pulsars is smaller than  the spatial resolution of HAWC. However, the tension between the VERITAS limits on B0355 and HAWC fluxes, as explained in Section~\ref{sec:res}, suggests that the TeV emission is more likely associated with J0359. 

\begin{deluxetable}{lccccc}
\tablecaption{Results of the likelihood fit assuming that the only emitting source is PSR B0355+54. The PL spectral model along with the two different spatial models were tested. \label{tab:b0355}}
\tablewidth{0pt}
\tablehead{
\colhead{Spatial model} & \colhead{TS} & \colhead{$\Delta{\rm BIC}$} & \colhead{Extension} & \colhead{$N_0$} & \colhead{$\alpha$}\\
\colhead{} & \colhead{} & \colhead{} & \colhead{[$\circ$]} & \colhead{$\rm TeV^{-1}~cm^{-2}~s^{-1}$} & \colhead{}
}
\startdata
Point-like & 35.86 & -1.9 & 0.0 & $(1.28_{-0.27}^{+0.34})\times 10^{-16}$ & $2.56 \pm 0.17$\\
Extended & 41.83 & -1.5 & $0.22 \pm0.09$ & $(2.0_{-0.5}^{+0.7})\times 10^{-16}$ & $ 2.51 \pm 0.15$\\
\enddata
\tablecomments{All associated errors are statistical. $\Delta{\rm BIC}$ is obtained comparing the BIC value with the best spectral model fit for both spatial models assuming that the emission is coming from J0359 (Section~\ref{sec:res}).}
\end{deluxetable}

\section{Fitting results of a two-source scenario}\label{app:both}
We further explore a scenario where both J0359 and B0359 contribute to the TeV emission observed by HAWC. Such a two-source model is disfavored by the data. 

Table \ref{tab:two-sources} presents the results of the two-source models. We consider three combinations of spatial profiles of the two sources: (A) both sources are point-like, (B) both sources are extended with a Gaussian shape, and (C) J0359 is extended source and B0355 is point-like. The energy spectrum is assumed to be a PL. The normalization flux $N_0$ and the spectral index $\alpha$ in each fit were free to vary while the position of the sources for all the scenarios were fixed. 

The $\Delta$TS column shows the gain of test statistics by adding an extra source to the one-source model presented in Section~\ref{sec:res} and Section~\ref{app:b0355} (the baseline model considers pure background plus the emission from the other source). The two-source model is disfavored in all cases. 

\begin{deluxetable}{llccccc}
\tablecaption{Results of the likelihood fit assuming that the excess observed comes from two sources: PSR J0359+5414 and PSR B0355+54. The spectral model for all the spatial models is a PL. \label{tab:two-sources}}
\tablewidth{0pt}
\tablehead{
\colhead{Two-source model} & \colhead{Source} & \colhead{$\Delta$TS}& \colhead{$\Delta{\rm BIC}$} & \colhead{Extension} & \colhead{$N_0$} & \colhead{$\alpha$}\\
\colhead{} & \colhead{} & \colhead{}&\colhead{} & \colhead{[$\circ$]} & \colhead{$\rm TeV^{-1}~cm^{-2}~s^{-1}$} & \colhead{}
}
\startdata
{    } & J0359 &2.32& {} & 0.0 & $(1.0_{-0.9}^{+0.5})\times 10^{-16}$ & $2.63_{-0.20}^{+0.6}$\\
Model A &B0355 &0.32& -24& 0.0 & $(0.00034_{-0.00024}^{+6})\times 10^{-13}$ & $2.4_{-5}^{+1.3}$ \\
\hline
{    } & J0359 &8.73& {     }& $1.500_{- 0.004}^{+0.18}$ & $(3.3_{-3.3}^{+1.5})\times 10^{-16}$ & $2.2_{-1.3}^{+0.4}$\\
Model B & B0355 &10.29& -26 & $0.14_{-0.15}^{+0.08}$ & $(1.5_{-0.4}^{+0.5})\times 10^{-16}$ & $2.56\pm0.20$ \\
\hline
{    } & J0359 &13.02& {     } & $1.5000\pm0.0010$ & $(0.04_{-0.04}^{+4})\times 10^{-14}$ & $2.2\pm2.8$\\
Model C & B0355 &8.62& -26 / -15 & 0.0 & $(3.2_{-1.5}^{+3.0})\times 10^{-16}$ & $2.60\pm0.28$
\enddata
\tablecomments{All associated errors are statistical. Model A corresponds to a scenario where both sources are point-like, model B assumes that both sources are extended with a Gaussian shape, and model C assumes that PSR J0359+5414 is as a point-like source and PSR B0355+54 is an extended source with a Gaussian shape. $\Delta{\rm BIC}$ is obtained comparing the BIC value with the best model fit assuming that the emission is coming from J0359 (Section~\ref{sec:res}). For model A, with the PL point-like model, for model B with the PL Gaussian model and for model C with the two previous models.}
\end{deluxetable}
\end{document}